\documentclass[nonacm, sigconf]{acmart}

\usepackage{tcolorbox}
\usepackage{array}
\usepackage{listings}

\usepackage{pifont}% http://ctan.org/pkg/pifont

\usepackage{xcolor}
\usepackage{soul}

\newcolumntype{L}[1]{>{\raggedright\let\newline\\\arraybackslash\hspace{0pt}}m{#1}}
\newcolumntype{C}[1]{>{\centering\let\newline\\\arraybackslash\hspace{0pt}}m{#1}}
\newcolumntype{R}[1]{>{\raggedleft\let\newline\\\arraybackslash\hspace{0pt}}m{#1}}

\newcommand{\codefont}[1]{\footnotesize{\texttt{#1}}\normalsize}
\newcommand{\totalproject}{35}

\AtBeginDocument{%
  \providecommand\BibTeX{{%
    \normalfont B\kern-0.5em{\scshape i\kern-0.25em b}\kern-0.8em\TeX}}}

\setcopyright{acmcopyright}
\copyrightyear{2019}
\acmYear{2019}
\acmDOI{10.1145/1122445.1122456}

\acmConference[ICSE '20]{ICSE '20: International Conference on Software Engineering}{May 23--29, 2020}{Seoul, South Korea}
\acmBooktitle{ICSE '20: International Conference on Software Engineering,
  May 23--29, 2020, Seoul, South Korea}
\acmPrice{15.00}
\acmISBN{978-1-4503-9999-9/18/06}

\hypersetup{draft}
\begin{document}
\title{Automatic Detection and Resolution of Software Merge Conflicts: Are We There Yet?}
%\title{An Empirical Study of Software Merge Conflicts and Their Resolutions}
%\title{A Characterization Study of Software Merge Conflicts}
%\title{Software Merge Conflicts: Challenges and Opportunities}
%\title{Detecting and Resolving Software Merge Conflicts: Challenges and Opportunities}


\author{Bowen Shen}
\email{bowenshe@vt.edu}
\affiliation{Virginia Polytechnic Institute and State University, USA}
\author{Cihan Xiao}
\email{cihan97@vt.edu}
\affiliation{Virginia Polytechnic Institute and State University, USA}
\author{Na Meng}
\email{nm8247@cs.vt.edu}
\affiliation{Virginia Polytechnic Institute and State University, USA}
\author{Fei He}
\email{hefei@mail.tsinghua.edu.cn}
\affiliation{Tsinghua University, China}

\begin{abstract}
%background
Developers create software branches for tentative feature addition and bug fixing, and periodically merge branches to release software with new features or repairing patches. When the program edits from different branches textually overlap (i.e., \emph{textual conflicts}), or the co-application of those edits lead to compilation or runtime errors (i.e., \emph{compiling or dynamic conflicts}), it is challenging and time-consuming for developers to eliminate merge conflicts. 
Prior studies examined %the popularity of merge conflicts and 
how conflicts were related to code smells or software development process; tools were built to find and solve conflicts. 
 However, some fundamental research questions are still not comprehensively explored, including 
(1) how conflicts were introduced, (2) how developers manually resolved conflicts, and (3) what conflicts cannot be handled by current tools. 

For this paper, we took a hybrid approach that combines automatic detection with manual inspection to reveal 204 merge conflicts and their resolutions in 15 open-source repositories. %in the version history of 15 open-source projects.
Our data analysis reveals three phenomena. First, compiling and dynamic conflicts are harder to detect, although current tools mainly focus on textual conflicts. Second, in the same merging context, developers usually resolved similar textual conflicts with similar strategies. Third, 
developers manually fixed most of the inspected compiling and dynamic conflicts 
by similarly editing the merged version as what they did for one of the branches. 
Our research reveals the challenges and opportunities for automatic detection and resolution of merge conflicts; it also sheds light on related areas like systematic program editing and change recommendation. 
%Because the conflicting edits from different branches may textually overlap or not, or may trigger syntactic or semantic errors, in this paper, we classified merge conflicts into three categories: \emph{textual conflicts}, \emph{syntactic conflicts}, and \emph{semantic conflicts}. 
%To characterize the conflicts that cannot be detected or resolved by existing tools, we took multiple automatic approaches and conducted manual analysis to (i) reveal conflicts and (ii) study how developers resolved those conflicts. Among the \todo{A} real conflicts gathered from the version history of \todo{B} open-source projects, we had \todo{C} textual conflicts, \todo{D} syntactic conflicts, and \todo{E} semantic conflicts. 
%developers resolved \todo{G\%} of syntactic conflicts by changing the identifies of used types, methods, or fields; (4) developers resolved \todo{H\%} semantic conflicts by repeating their refactorings that were applied to one branch in the merged version.
%These observations deepen our understanding of software merge conflicts, and reveal the strategies developers frequently took to resolve conflicts. 

\end{abstract}



%Continuous Integration (CI)}

\maketitle

\section{Introduction}

%Software \emph{branch} and \emph{merge} are two fundamental aspects of version control systems~\cite{Pilato:2004:VCS}. As developers collaboratively work on a software product, a developer Alex may create a branch out of the trunk (``main line'') of software development for feature addition and bug fixing. Afterwards, Alex may also merge the branch with the trunk now and then to integrate his/her program changes with other developers' code. 

``Integration Hell'' refers to the scenarios when
developers integrate or \emph{merge} a big chunk of code changes at the last minute before delivering a software product~\cite{integrationhell}. 
In traditional software development environments, this integration process is rarely smooth and seamless, but results in \emph{conflicts}, which can take developers hours or perhaps days in fixing the code so that it can finally integrate~\cite{fastestUpdates}. 
To avoid ``Integration Hell'', more and more developers have recently adopted Continuous Integration (CI) to integrate code more frequently (e.g., several times a day), and to verify each integration via automated builds (i.e., compilation and testing)~\cite{Vasilescu:2015,Savor:2016}.  

Nevertheless, CI practices do not eliminate the challenges posed by merge conflicts. %In the CI process, 
Instead, developers
still mainly rely on the merge feature 
%When developers heavily rely on the merge functionality 
of version control systems (e.g., git-merge~\cite{git-merge}) to automatically (1) integrate branches and (2) reveal any conflict for manual resolution. However, \emph{such text-based merge usually produces lots of false positives and false negatives}. For example, when two branches reformat the same line in divergent ways (e.g., add vs.~ delete a whitespace), git-merge simply reports a \emph{textual conflict} although such conflicts are unimportant and cause no difference syntactically or semantically. Meanwhile, when two branches edit different lines and modify the program semantics or logic in conflicting ways, git-merge silently applies both edits with no conflict reported. 

%without reporting any conflict.
Tools were proposed to improve over text-based merge
~\cite{Brun:2011,Apel:2011,Apel:2012,Leenich2014,Zhu:2018}. Specifically, FSTMerge reduces false positives by modeling Java program entities (e.g., classes, methods, and fields) as unordered leaf nodes, and resolving textual conflicts when two edits insert unrelated declarations at the same location~\cite{Apel:2011}. 
JDime removes fake conflicts by modeling both program entities and statements in its tree representation, and by applying tree matching and amalgamation algorithms to resolve conflicts~\cite{Apel:2012,Leenich2014}. 
Given textually conflicting edits, AutoMerge uses Version Space Algebra (VSA) to enumerate all possible combinations of the edit operations
from both sides and to recommend alternative resolutions~\cite{Zhu:2018}. However,
\emph{none of these tools help reduce the false negatives of text-based merge}. 
Additionally, Crystal proactively monitors for developers' program commits in separate branches~\cite{Brun:2011}. By tentatively merging the latest commits between branches and building the merged software, Crystal notifies developers of 
\emph{higher-order} conflicts, i.e., the changes that are semantically incompatible. 

\begin{figure*}[h]
\centering
\includegraphics[width=0.75\textwidth]{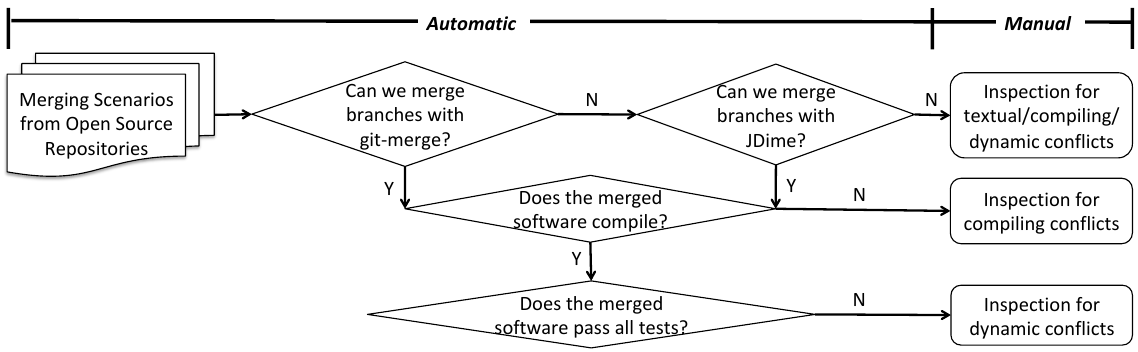}
\vspace{-0.5em}
\caption{Workflow of our hybrid approach to study merge conflicts and their resolutions}
\label{fig:workflow}
\end{figure*}

Despite existing tool support, our knowledge of software merge conflicts is still limited. For instance, although some tools (e.g., JDime and AutoMerge) refine and resolve the conflicts reported by text-based merge, it is still unknown whether these tools can 
%(1) eliminate all false alarms and (2) 
resolve all true conflicts fully automatically.  
%if these techniques are insufficient, what is the missing tool support. 
Even though Crystal can reveal different conflicts from git-merge, it is still unknown whether Crystal can capture all the conflicts missed by git-merge. 
\emph{Having a better understanding of merge conflicts and their resolutions is crucially important for two reasons.} 
First, by exploring the gap between challenges of merge conflicts and capabilities of existing tools, we can shed light on future tools to better aid developers. Second, by characterizing the limit of automatic tool support, we can design better human-in-the-loop approaches to focus developers' manual effort on the most important and challenging conflicts. 

For this paper, we conducted a comprehensive in-depth investigation on software merge conflicts and their resolutions. Because we are unsure whether existing tools can detect all kinds of conflicts, we took a hybrid approach that combines automatic tools with manual inspection. Specifically, as shown in Figure~\ref{fig:workflow}, we first applied both git-merge and JDime to the merging scenarios in \totalproject{} open-source repositories. 
Given a scenario with two branches, 
%branch pair, 
if neither tool could merge the branches, 
we examined related edits to identify various conflicts with our best effort. 
Otherwise, if the given pair is automatically mergeable, we further used automatic compilation and testing to identify any higher-order conflict. 
With this approach, we collected 100 textual conflicts, 100 compiling conflicts (i.e., incompatible changes causing compilation errors), and 4 dynamic conflicts (i.e., incompatible changes triggering abnormal program behaviors). We contrasted these conflicts with the capabilities of existing tools, and characterized the root cause and resolution of each conflict quantitatively and qualitatively.

By analyzing many conflicts unrevealed before, our research uncovers many interesting findings that have not been previously reported. 
%novel challenges and opportunities for automatic detection and resolution of merge conflicts. 
The major findings are summarized as follows. 
\begin{itemize}

\item \textbf{How were conflicts introduced?} 51\% of textual conflicts were caused by the contradictory statement updates between branches. 93\% of compiling conflicts occurred when one branch adds one or more references to a program entity that is updated, removed, or replaced in another branch. 75\% of dynamic conflicts happened because the test oracle added by one branch does not correspond to the code implementation updated by the other branch. 

\item \textbf{How did developers manually resolve such conflicts?} 
For 86\% of textual conflicts, developers resolved conflicts by (1) keeping the changes from one branch or (2) combining part or all of the edits from both sides. 
For compiling and dynamic conflicts, developers never purely combine edits from both branches; instead, they applied extra edits to the merged software such that all similar code locations were modified consistently.

\item \textbf{What conflicts cannot be handled by current tools?} 
In our data set, 79\% of compiling conflicts and 75\% of dynamic conflicts were not reported or reflected by any explored automatic approach, let alone to be resolved automatically. 
Although some tools could suggest resolutions for 92\% textual conflicts and 25\% dynamic conflicts, they rarely guarantee the correctness of their suggestions. 
%88\% of the inspected conflicts cannot be handled in this way. 

\end{itemize}
\emph{Importantly,  
we found that
although many conflicts cannot be detected or resolved by existing tools, the conflicts were introduced for typical reasons and 
developers took certain ways to resolve those conflicts. }
%By characterizing such conflicts in various aspects, 
Our study will enlighten future software merge tools, and suggest future research directions in related areas like automatic program transformation and change recommendation. 

\vspace{-0.5em}
\section{Background}
This section defines the terminology used in this paper (Section~\ref{sec:terms}), and  introduces the software merge tools we explored (Section~\ref{sec:tools}). 

\vspace{-0.5em}
\subsection{Terminology}
\label{sec:terms}
When developers merge two branches (e.g., $b_l$ and $b_r$) in a software repository,  there can be three types of conflicts.  

\begin{figure}[h]
\centering
\includegraphics[width=0.36\textwidth]{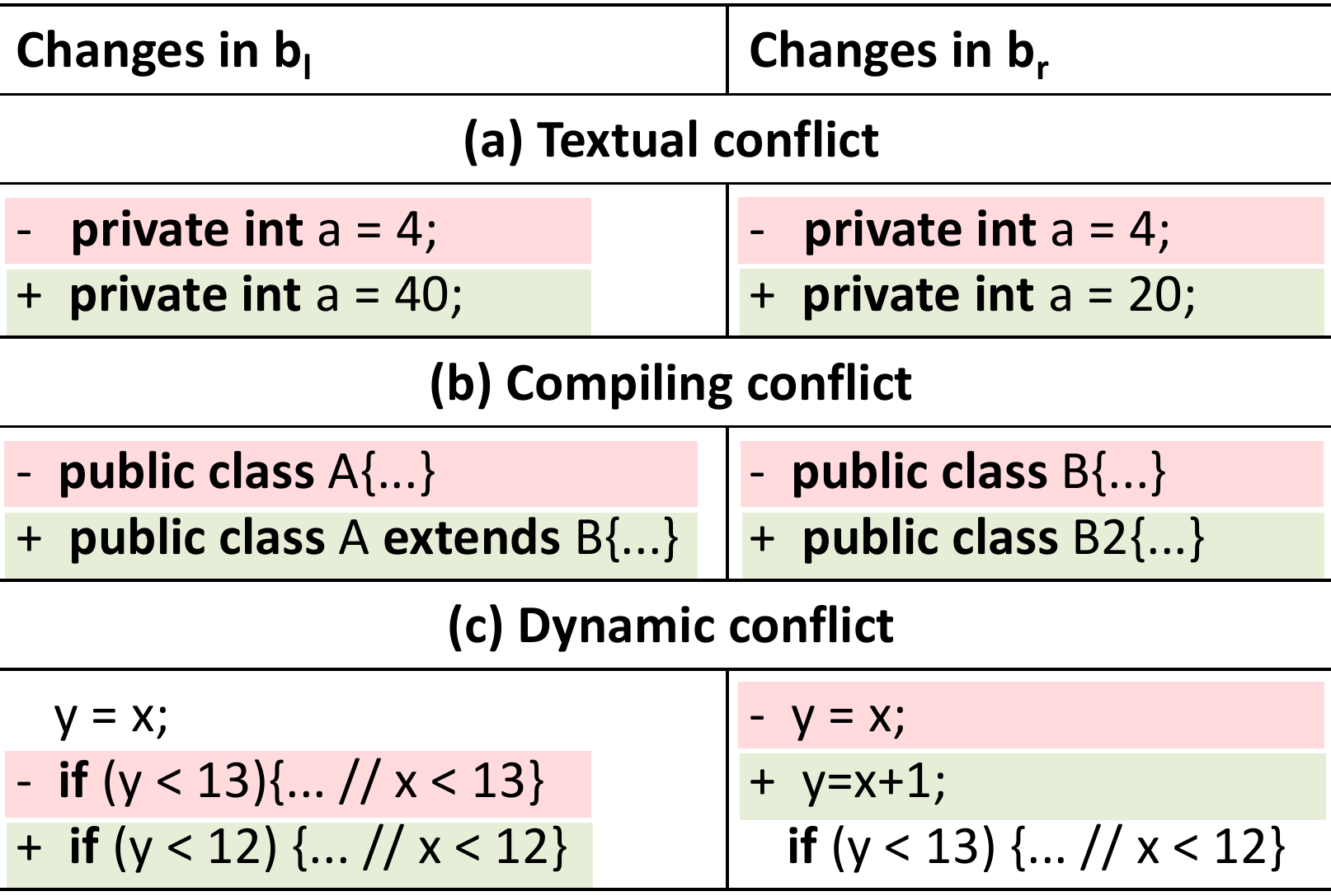}
\vspace{-0.5em}
\caption{Exemplar merge conflicts}
\label{fig:conflicts}
\end{figure}
%\noindent
\textbf{1. Textual Conflicts} exist when $b_l$ and $b_r$ edit the same line of text. As illustrated by Figure~\ref{fig:conflicts} (a), since $b_l$ and $b_r$ update the same statement with conflicting values (e.g., \codefont{40 vs.~20}), there is a textual conflict between the branches.  

\textbf{2. Compiling Conflicts} happen when (1) the edits of $b_l$ and $b_r$ do not have any textual conflict, and (2) the co-application of both edits triggers a \emph{compilation} error. As shown in Figure~\ref{fig:conflicts} (b), when $b_l$ adds a reference to class \codefont{B} and $b_r$ renames \codefont{B} to \codefont{B2}, the integration of these edits can make the \codefont{B}-reference unresolvable. 

\begin{figure*}
\includegraphics[width=0.8\textwidth]{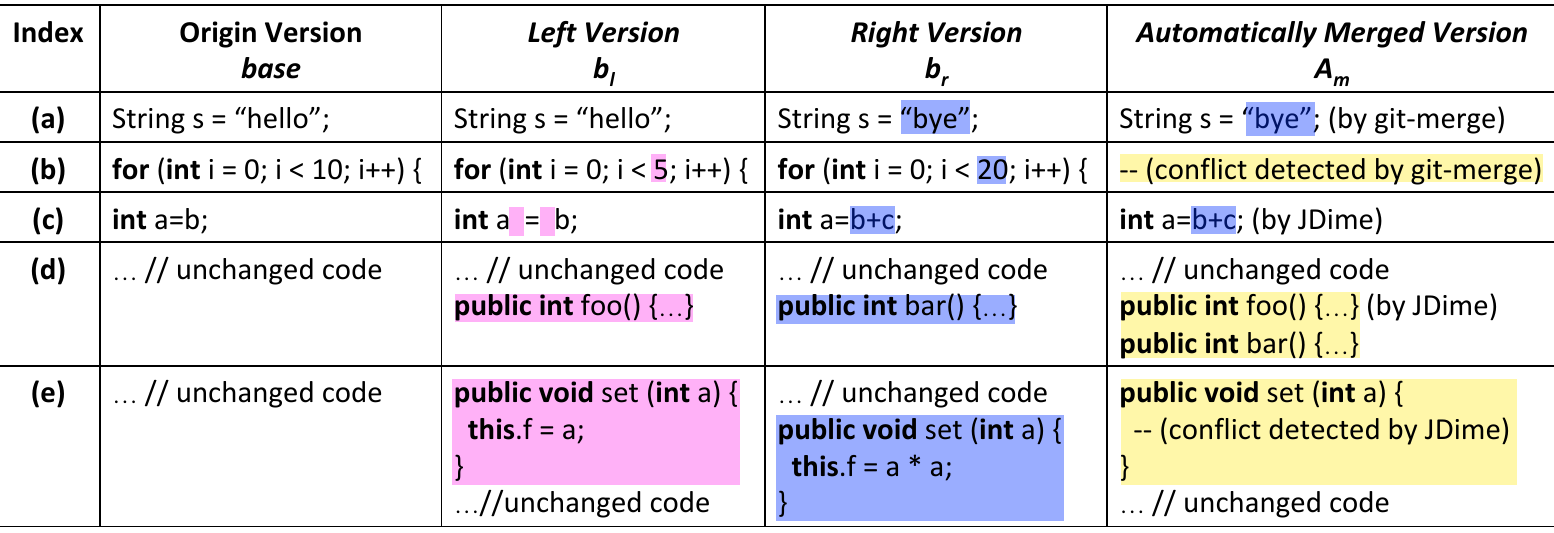}
\vspace{-0.5em}
\caption{Merging scenarios where $b_l$ and $b_r$ are different}
\label{fig:background}
\end{figure*}

\textbf{3. Dynamic Conflicts} occur when (1) the edits of $b_l$ and $b_r$ do not have any textual conflict, and (2) the co-application of both edits triggers a \emph{runtime error} or \emph{unexpected program behavior}. For example, in Figure~\ref{fig:conflicts} (c), although the edits in $b_l$ and $b_r$ separately satisfy the invariant ``\codefont{x < 12}'' inside the \codefont{then}-branch, applying both of them dissatisfies the invariant. 

\vspace{-0.5em}
\subsection{Software Merge Tools We Explored}
\label{sec:tools} 
In our research, we directly adopted text-based merge (i.e., git-merge) and JDime to detect textual conflicts. We also mimicked the workflow of Crystal to reveal compiling and dynamic conflicts. 

\subsubsection{Text-Based Merge (also called ``unstructured merge''~\cite{Apel:2011})}
It is the default merge feature provided by various version control systems (e.g., Git and SVN).
We used ``git-merge'' in our study, which conducts a \textbf{three-way merge}~\cite{Mens2002} to analyze and resolve differences between branches. 
Suppose that $b_l$ and $b_r$ derive from the same origin---$base$. Whenever $b_l$ and $b_r$ have different text for the same line of code, three-way merge further compares both branches with $base$ to decide which branch changed the line. 
As shown in Figure~\ref{fig:background} (a), if only one branch (i.e., $b_r$) changes the line, the change is kept in the automatically merged software $A_m$; otherwise, if both branches modify the line in divergent ways (see Figure~\ref{fig:background} (b)), three-way merge reports a textual conflict.  

\subsubsection{JDime (also called ``structured merge''~\cite{Apel:2012})}
%It merges software by observing program syntactic structures. 
Given a Java file, JDime creates a syntax tree by modeling declarations of program entities (imports, classes, methods, and fields) as \emph{unordered nodes}, and modeling Java statements as \emph{ordered tree nodes}.
%With its tree matching and amalgamation algorithms, 
 JDime matches unordered nodes based on the content and subtrees of each node; it matches ordered nodes by also considering the sequential ordering between nodes. 
\emph{Intuitively, given a textual conflict $C$, if JDime can resolve $C$ via structured merge, it does not report the conflict; otherwise, JDime still reports $C$}. 

Theoretically, JDime can improve over git-merge in three ways. First, as JDime compares syntax trees to infer changes, it can suppress the unimportant conflicts due to formatting changes. As shown in Figure~\ref{fig:background} (c), when $b_l$ inserts whitespaces and $b_r$ modifies the assignment logic, JDime simply keeps the logic change in $A_m$ instead of reporting any conflict. 
Second, JDime compares entity declarations by ignoring their code locations. Therefore, when $b_l$ and $b_r$ insert distinct declarations at the same code location (as illustrated in Figure~\ref{fig:background} (d)), JDime can resolve the textual conflict by inserting both declarations to $A_m$. Third, JDime can reveal some conflicts missed by git-merge. For instance, when $b_l$ and $b_r$ insert same-named entity declarations at different code locations (see Figure~\ref{fig:background} (e)), JDime is able to correlate these insertions and decide whether the declarations contain any textual conflict. 

\subsubsection{Crystal~\cite{Brun:2011}}
When developers work on different software branches and commit program changes now and then, Crystal speculatively merges the latest versions between branches and notifies developers of potential conflicts, before developers conduct any actual merge. 
As Crystal is an interactive tool that works while developers actively work on different software branches, it does not fit into our history-based merging scenario analysis. Therefore, we did not run Crystal, but followed Crystal's three-step strategy to reveal different kinds of conflicts:
\begin{itemize}
\item \textbf{Step 1}: If $b_l$ and $b_r$ cannot be integrated via text-based merge, there is one or more textual conflicts. 
\item \textbf{Step 2}: If $b_l$ and $b_r$ compile and can be merged automatically, but the merged software $A_m$ fails to compile,  
%\hl{whereas both $b_l$ and $b_r$ can compile}, 
then there is at least one compiling conflict. 
\item \textbf{Step 3}: If $b_l$ and $b_r$ pass their separate test suites and can be merged automatically, but the merged software $A_m$ fails its test suite,  
%\hl{whereas both $b_l$ and $b_r$ can pass its test suite respectively}, 
then there is at least one dynamic conflict. 

\end{itemize}

Our goal of trying these tools is not to evaluate their implementation status. Instead, we focus on the limitation of their approach design. 
Assuming that existing methodologies are perfectly implemented, we were curious 
what are the conflicts that have not be automatically handled yet and what is the missing tool support.

\vspace{-0.5em}
\section{Study Approach}
\label{sec:approach}

As shown in Figure~\ref{fig:workflow}, 
our approach has two phases. Phase I uses automatic approaches to detect conflicts or identify errors due to conflicts (Section~\ref{sec:automatic}). Phase II adopts manual inspection to examine the merging commits reported by Phase I or reveal extra conflicts, creating a data set of conflicts and their resolutions (Section~\ref{sec:manual}). 

\vspace{-0.5em}
\subsection{Phase I: Automatic Detection of Conflicts or Their Symptoms}
\label{sec:automatic}

\begin{figure}[h]
\includegraphics[width=0.3\textwidth]{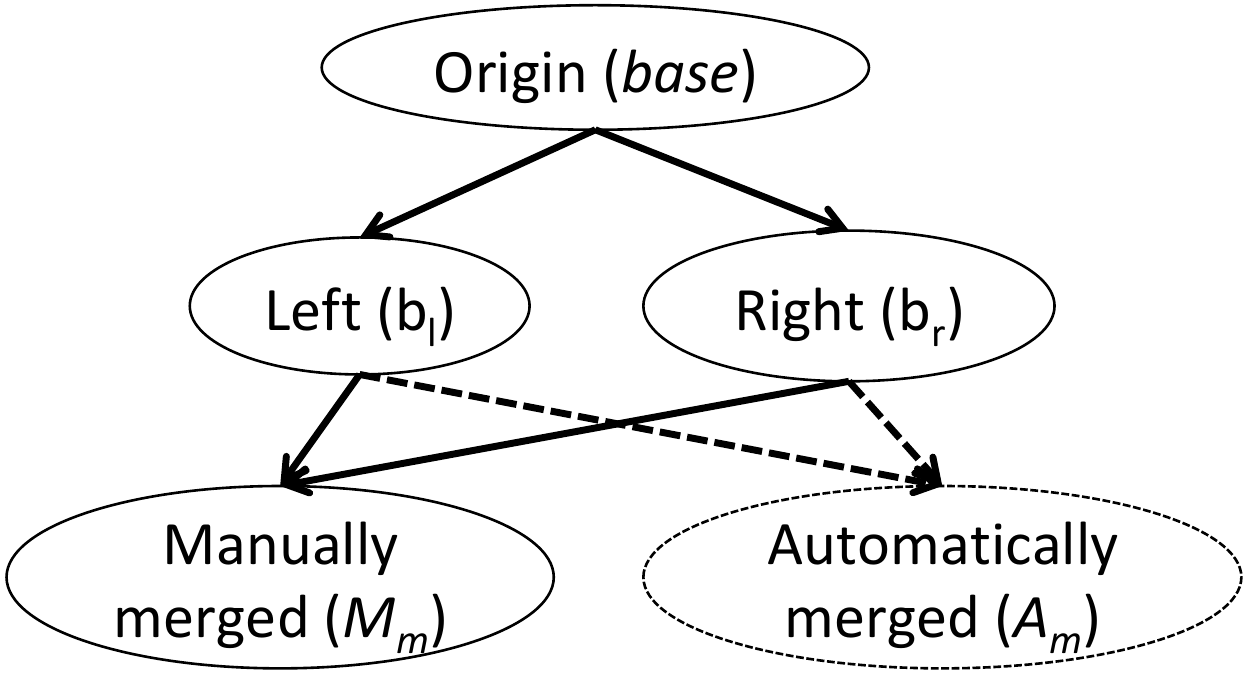}
%\vspace{-0.5em}
\caption{Software versions related to a merging scenario}
\label{fig:scenario}
\end{figure}

Given a software repository, for each merging commit or scenario (see Figure~\ref{fig:scenario}), 
there are two branches to merge---left ($b_l$) and right($b_r$), the most recent common ancestor of those branches ($base$), and the merged version created by developers ($M_m$).  
%and an optional merged version generated by tools ($A_m$). 
With all merging scenarios identified in the version history, similar to Crystal, our approach took three steps to reveal conflicts. 

In Step 1, we first applied git-merge to $b_l$ and $b_r$, in order to generate an automatically merged version $A_m$ (see Figure~\ref{fig:scenario}). If git-merge failed to produce $A_m$ due to textual conflicts in some Java files, we further applied JDime especially to those files, hoping to generate $A_m$ successfully.   
We intentionally applied the tools in this specific order for three reasons. 
First, git-merge compares branches faster while JDime better resolves conflicting Java edits. By focusing JDime on the textual conflicts that git-merge cannot process, our approach can efficiently detect true textual conflicts. 
%resolve as many textual conflicts as possible. 
Second, git-merge can propagate file-level operations
(e.g., moving files across folders) to $A_m$, while JDime cannot. 
By using both tools, 
%only examines Java file updates.  
%With git-merge used to propagate file-level operations to $A_m$, 
we ensure that the merged program is more likely to compile than the one generated solely by JDime. 
Third, we intended to reveal any conflicts unsolvable by existing tools, so JDime was used as a filter to refine the textual conflicts reported by git-merge.

In Step 2, we further compiled 
$b_l$, $b_r$ and the $A_m$ produced by Step 1. 
If both branches compile successfully but $A_m$ fails to compile, we conclude that 
the branches have at least one compiling conflict; otherwise, if either branch does not compile, we skip the commit. 
Currently, our research handles the programs compilable by Maven~\cite{maven}, Ant~\cite{ant}, or Gradle~\cite{gradle}.
% For example, given a Maven project, our approach leverages the command ``mvn clean compile'' to compile the program into executable software. The output of this step is either a compiled program or any error reported during compilation. 
Notice that this step cannot directly pinpoint the location of any compiling conflict; instead, it presents symptoms caused by those conflicts. 

In Step 3, we ran the compiled versions of
$b_l$, $b_r$ and $A_m$ with their separate test suites. If both branches pass all tests while $A_m$ fails any test, our approach concludes that the branches have at least one dynamic conflict. 
%$A_m$ contains at least one semantic error. 
%For instance, given a Gradle project, our approach leverages the command ``./gradlew test'' to trigger test execution.
%If $A_m$ cannot pass all tests, this step outputs the test failure(s). 
Similar to Step 2, this step presents effects of dynamic conflicts instead of showing the conflicts themselves. 

\vspace{-0.5em}
\subsection{Phase II: Manual Inspection of Conflicts and Their Resolutions}
\label{sec:manual}

The ultimate goal of our manual inspection is to 
%When examining the outputs by Phase I, our manual inspection intends to 
(1) reveal as many true conflicts as possible and (2) remove the false conflicts reported by tools.
However, for individual types of conflicts, our inspection serves slightly different purposes. 

\subsubsection{Refinement of Detected Textual Conflicts}
Given a textual conflict reported by Step 1, we first determined whether the conflict is a false positive. Due to some implementation issues, JDime sometime failed to eliminate the textual conflicts that its methodology is supposed to handle. Again, our study does not examine the limitation of existing tool implementation. Instead, it explores the conflicts overlooked
by current automatic approaches. Therefore, we manually filtered out the textual conflicts that should have been resolved by a perfect implementation of JDime.
Additionally, since it is almost infeasible to include all true (i.e., unresolved) textual conflicts into our study, we gathered 100 samples of such conflicts for further analysis.  
Especially, for each gathered sample, we compared $b_l$, $b_r$, and the manually merged version $M_m$ to learn (1) how the conflict was introduced and (2) how developers resolved it. 

\subsubsection{Diagnosis for Compilation Errors and Test Failures}

Given an error reported by Step 2 or Step 3, we identified the conflict by examining the error message, source code, and related edits from both branches. 
Because test failures are usually harder to reason about than compilation errors, 
 our manual analysis does not guarantee to reveal the root cause for any observed  test failure. 
 Additionally, for each identified conflict, 
we also inspected the corresponding resolution edits applied by developers in $M_m$.

\subsubsection{Manual Detection of Higher-Order Conflicts}
Based on our experience, Step 2 and Step 3 did not generate many errors or failures for further examination.
To collect more data, we also inspected the unmerged branches in Step 1 to recognize more conflicts. 

\begin{figure}[h]
\centering
\includegraphics[width=0.27\textwidth]{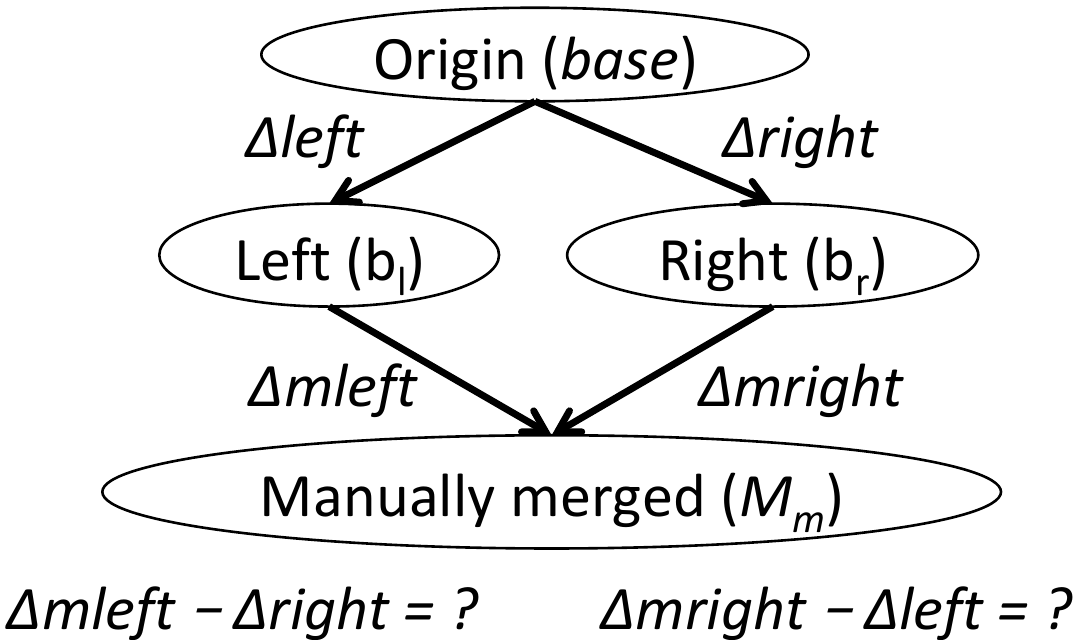}
\vspace{-0.5em}
\caption{Manual detection of compiling/dynamic conflicts}
\label{fig:manual-conflict}
\end{figure}

Our manual process is shown in Figure~\ref{fig:manual-conflict}. 
Given a merging scenario where branches cannot be automatically merged, we identify the edits that 
 developers applied in individual branches as $\Delta left$ and $\Delta right$. We also denote the difference between the left branch and the merged version as $\Delta mleft$, and the difference between the right branch and the merged version as $\Delta mright$.
Formally,  
 \begin{align}
     M_m &= apply(apply(base, \Delta left), \Delta mleft) \nonumber\\
         &= apply(apply(base, \Delta right), \Delta mright) \nonumber
 \end{align}
Ideally, if the two branches can be automatically integrated without any conflict, $\Delta mleft = \Delta right$ and $\Delta mright = \Delta left$. However, for the scenarios where branches cannot be automatically merged, by comparing $\Delta right$ with $\Delta mleft$ (or comparing $\Delta left$ with $\Delta mright$), we can learn how developers manually resolved conflicts. 
For each observed difference, we speculated whether any compilation or runtime error would occur if developers had not made the observed edit. 
Since we do not have sufficient knowledge for any project under study, our speculative analysis is limited to revealing obvious compiling or dynamic conflicts. 

For one commit of {\sf elasticsearch}~\cite{merge-example}, 
we compared the diff files and 
found the following difference between $\Delta right$ and $\Delta mleft$:

\begin{lstlisting}
< +   if (parseContext.queryTypes().size() > 1) {
---
> +   if (parseContext.shardContext().queryTypes().size() > 1) {
\end{lstlisting}
\noindent
It means that to merge the branches, developers manually updated a method call (i.e., \codefont{parseContext.queryTypes()}), which was added in $\Delta right$. To confirm that this update is mandatory for software merge, 
%decide whether this update is %required by software merge, 
we further checked the edits in $\Delta left$, finding 
that the related method declaration was removed. This implies that without the manual update, the two branches have at least one compiling conflict because $b_r$ added a reference to a method removed by $b_l$.

For this phase, we have three authors involved in manual inspection. For each conflict included in our data set, we ensured that at least two people examined and described the merging scenario and conflict resolution. When unsure about certain conflicts, we had discussions to achieve consensus. 

\vspace{-0.5em}
\section{Evaluation}
This section first introduces the open source projects studied and the data set constructed (Section~\ref{sec:data}); 
it then explains the experiment design and findings for each research question (Section~\ref{sec:rq1}-\ref{sec:rq3}).

\begin{comment}
\begin{table}
\centering
\footnotesize
\caption{Subject projects}
\label{tab:data}
\vspace{-0.5em}
\begin{tabular}{l| R{0.8cm} R{0.8cm} R{1.2cm} R{1.5cm}}
\toprule
\textbf{Project Name} & \textbf{\# of KLOC} & \textbf{\# of Stars} & \textbf{\# of Branches} & \textbf{\# of Merging Scenarios}\\ \toprule
Activiti \\ \hline
bigbluebutton \\ \hline
buildergenerator \\ \hline
cassandra \\ \hline
caelum-stella \\ \hline
checkstyle\\ \hline
closure-compiler \\ \hline
coveralls-maven-plugin \\ \hline
elasticsearch \\ \hline
error-prone \\ \hline
fastjson \\ \hline
hamcrest-bean \\ \hline
htmlelements \\ \hline
httpcomponents-core \\ \hline
HikariCP\\ \hline
HdrHistogram \\ \hline
jade4j \\ \hline
javapoet \\ \hline
jongo \\ \hline
lombok \\ \hline
mybatis-3 \\ \hline
named-regexp \\ \hline
nutz \\ \hline
nuxeo \\ \hline
orientdb \\ \hline
pebble \\ \hline
pmd \\ \hline
raml-java-parser \\ \hline
spoon \\ \hline
spring-hateoas \\ \hline
truth \\ \hline
vectorz \\ \hline
webmagic \\ \hline 
wildfly \\
\bottomrule
\end{tabular}
\end{table}
\end{comment}

\vspace{-0.5em}
\subsection{Data Set Construction}
\label{sec:data}
We experimented with the repositories of \totalproject{} open source projects.
%For each project, 
% Table~\ref{tab:data} presents \textbf{\# of KLOC} (i.e., code size), \textbf{\# of Stars} (i.e., popularity), \textbf{\# of Branches}, and \textbf{\# of Merging Scenarios}.
33 of the projects were included because they were mentioned in prior work ~\cite{Apel:2012,long2017automatic}. Besides, 
we also included 
{\sf nuxeo}~\cite{nuxeo} and {\sf elasticsearch}~\cite{elasticsearch} because  these projects are large and popular, containing many branches and merging scenarios. Table~\ref{tab:distribution} shows the conflicts we found in these projects and included in our data set. 
As mentioned in Section~\ref{sec:approach}, there are so many real textual conflicts revealed by tools that it is infeasible to manually analyze all of them. Therefore, we sampled 100 of the revealed conflicts in 7 projects. 

\begin{table}
\centering
\footnotesize
\caption{Observed conflicts included in our sample data}
\label{tab:distribution}
\vspace{-0.5em}
\begin{tabular}{l| R{1.5cm} R{1.8cm} R{1.5cm}}
\toprule
\textbf{Project Name} & \textbf{\# of Textual Conflicts} & \textbf{\# of Compiling Conflicts} & \textbf{\# of Dynamic Conflicts}\\ \toprule
orientdb & 32 & - & - \\ \hline
wildfly & 7 & *3 & -\\ \hline
pmd & 13 & - & -\\ \hline
lombok & 13 & - & - \\ \hline
bigbluebutton &7 & - & - \\ \hline
cassandra & 15 & - & -\\ \hline
Activiti & 13 & 8 & -\\ 
&  & *1 & \\ \hline
fastjson & - & 2 & -\\  \hline
javapoet & - & 1 & - \\ \hline
pebble & - & 4 & 1 \\ \hline
truth & - & 2 & - \\ \hline
vectorz & - & 2 & - \\ \hline
webmagic & - & 2 & - \\ \hline
nuxeo & - & *1 & - \\ \hline
elasticsearch & - & *74 & *3\\ 
\bottomrule
\textbf{Sum} & \textbf{100} & \textbf{100} & \textbf{4}\\ \bottomrule
\multicolumn{4}{l}{``-'' means no conflict of certain type(s) is revealed or included in the data set.}\\
\multicolumn{4}{l}{``*'' implies that the reported conflicts were manually detected.}
\end{tabular}
\end{table}

Based on the compilation of automatically merged software ($A_m$), we only identified 21 compiling conflicts in 7 projects; among 5 of these projects, automatic compilation could reveal at most 2 conflicts for each project. 
In contrast, we manually inspected the automatically unmergeable branches in four projects--- {\sf Activiti}, {\sf wildfly}, {\sf nuxeo}, and {\sf elasticsearch}---and detected many more conflicts. 
For simplicity, we included 79 of the manually found conflicts, obtaining 100 compiling conflicts in total.  
%such that our data set contains 100 syntactic conflicts in total. 

We identified only four dynamic conflicts in total, one of which was revealed by automatic compilation and testing, while three were manually detected. 
Three reasons can explain the small number of obtained conflicts. First, 
when both $b_l$ and $b_r$ can pass their separate test suites and $A_m$ can compile,  
%when the automatically merged software ($A_m$) can compile, 
$A_m$ always passes all tests. Second, when $A_m$ fails a test, it is quite challenging to reason about the root cause. We actually had three more merging scenarios where $A_m$ failed at least one test. However, since we could not understand how those test failures were related to software merge, we did not include them into our data set. Third, manually detecting dynamic conflicts is also very time-consuming and error-prone. Although we inspected the automatically unmergeable branches and developers' resolutions in one  project---{\sf elasticsearch}---we could rarely 
assess what runtime errors would occur had developers not applied their edits in the merging commits. 

\vspace{-0.5em}
\subsection{RQ1: How Were Conflicts Introduced?}
\label{sec:rq1}

This section introduces our characterization for different kinds of conflicts (Section~\ref{sec:tcIntro}-Section~\ref{sec:s2cIntro}).

\subsubsection{Characterization of Textual Conflicts}
\label{sec:tcIntro}

As shown in Table~\ref{tab:textualCause}, we identified six reasons to explain how textual conflicts were introduced. Specifically, 51 conflicts happened when $b_l$ and $b_r$ updated the same statement(s) in distinct ways. 
29 conflicts occurred because a branch deleted certain statement(s) while the other branch updated the same statement(s). 15 conflicts were introduced because of conflicting statement insertion. For instance, suppose that $b_l$ inserts the following \codefont{if}-statement:
\begin{lstlisting}
if (!genericArgs.isEmpty()) {
  sb.delete(sb.length() - 3, sb.length() - 1);
}
\end{lstlisting}
while $b_r$ inserts a similar \codefont{if}-statement at the same location:
\begin{lstlisting}
if (!genericArgs.isEmpty()) {
  sb.replace(sb.length() - 3, sb.length() - 1, "");
}
\end{lstlisting}
The two statements are different because $b_l$ invokes \codefont{StringBuffer.delete (...)} method to remove some characters, while $b_l$ invokes a different method \codefont{StringBuffer.replace(...)} to achieve the same goal. 
%As shown in Figure~\ref{fig:textualCause} (c), both branches insert an \codefont{if}-statement. However, $b_l$ invokes \codefont{StringBuffer.delete(...)} method to remove some characters, while $b_l$ invokes a different method \codefont{StringBuffer.replace(...)} to achieve the same goal. 

\begin{table}[h]
%\caption{Scenarios where textual conflicts were introduced}
%\caption{Six ways in which textual conflicts were introduced}
\caption{Classification of textual conflicts}
\label{tab:textualCause}
%\vspace{-0.5em}
\footnotesize
\begin{tabular}{llr}
\toprule
\textbf{Idx} & \textbf{Conflict Type} & \textbf{\# of Conflicts} \\ \hline
(a) & \textbf{update-update} & 51 \\ \hline
(b) & \textbf{delete-update} & 25 \\ \hline
(c) & \textbf{insert-insert} & 15 \\ \hline
(d) & \textbf{update-move} & 6 \\ \hline
(e) & \textbf{delete-move} & 2 \\ \hline
(f) & \textbf{move-move} & 1 \\ 
\bottomrule
\end{tabular}
\end{table}

\begin{table*}
\footnotesize
%\caption{Scenarios where compiling conflicts were introduced}
\caption{Classification of compiling conflicts based on their introduction}
\label{tab:scIntro}
\vspace{-0.5em}
\begin{tabular}{L{1.2cm}| L{4cm}| L{10.3cm}| R{0.8cm}}
\toprule
\textbf{Entity} & \textbf{Conflict Type} & \textbf{Description} & \textbf{\# of Conflicts}\\
\toprule
Class- & Referencing a missing class $C$ & One branch adds a reference to $C$, 
while the other branch renames, replaces, or removes $C$, or removes the import declaration of $C$. & 17
\\ \cline{2-4}
related& Importing a missing class $C$ & One branch imports a class declaration of $C$, while the other branch removes the dependency library that declares $C$. & 1 \\ \hline
Interface- & Referencing a missing interface $I$ & One branch adds a reference to $I$, while the other branch renames $I$. & 1 \\ \cline{2-4}
related& Restructuring an interface $I$ & One branch declares a new method in $I$, while the other branch redefines the interface as an abstract class. & 2\\ \hline
Enum-related & Referencing a missing enum data type $E$ & One branch adds a reference to $E$, while the other branch remove $E$ or move it into a different Java class. & 2 \\ \hline 
 & Referencing a missing method $M$ & One branch adds an invocation to $M$, while the other branch (1) renames, replaces, removes $M$, (2) removes the import declaration of $M$, or changes the parameter list. & 58 \\ \cline{2-4} 
Method-& Invoking a method $M$ with an updated return type & One branch adds an invocation to $M$, while the other branch changes the return type of $M$. & 2 \\ \cline{2-4}
related& Updating the parameter list of $M$ & One branch updates one parameter, while the other branch adds a new parameter. & 1\\ \cline{2-4} 
 & Restructuring a method $M$ & One branch changes the parameter list, while the other branch overrides $M$. & 2 \\ \hline
 Field-related & Referencing a missing field $F$ &One branch adds a reference to $F$, while the other branch renames or replaces $F$. & 12 \\ \hline
Var- & Referencing a missing variable $V$ & One branch adds a reference to $V$, while the other branch removes $V$. & 1 \\ \cline{2-4}
related& Adding duplicated declarations for variable $V$ & Two branches separately add a declaration of $V$ at different locations. & 1 \\  
\bottomrule
\end{tabular}
\end{table*}

Additionally, 6 conflicts occurred because a branch updated a statement while the other branch moved the statement;  
%For instance, Figure~\ref{fig:textualCause} (d), $b_l$ moves the \codefont{filled}-declaration statement into a \codefont{try}-clause while $b_r$ updates the statement. 
2 conflicts happened because one branch deleted a statement while the other branch moved the statement. 
%As shown in Figure~\ref{fig:textualCause} (e), $b_l$ deletes an \codefont{if}-statement, while $b_r$ moves the statement into a \codefont{try}-clause. 
Finally, one conflict was introduced because both branches moved the same statement to different places.  
%In Figure~\ref{fig:textualCause} (f), $b_l$ moves a statement into the \codefont{then}-block of a newly inserted \codefont{if}-statement. Meanwhile, $b_r$ moves the same statement into the \codefont{then}-block of another newly inserted \codefont{if}-statement. 

\begin{tcolorbox}
	\textbf{Finding 1:} \emph{
	66\% of textual conflicts happened when branches applied \textbf{conflicting updates or insertions}; the remaining textual conflicts occurred because one branch \textbf{updated or moved} certain statements, while the other branch \textbf{deleted} those statements or \textbf{moved} the statements to a different place.
}
\end{tcolorbox}

\subsubsection{Characterization of Compiling Conflicts}
\label{sec:scIntro}

We classified the observed compiling conflicts based on two factors: (1) the major entities involved, and (2) the edits producing the conflicts. As shown in Table~\ref{tab:scIntro}, each conflict involves one of the following six program entities: classes, interfaces, enums, methods, fields, and variables. In particular, 63\% of conflicts were related to methods, 18\% of conflicts were relevant to classes, and 12\% of conflicts were about field usage. 
Specifically among the method-related conflicts, 58 conflicts were introduced because a branch added a reference (invocation) to method $M$, and the other branch voided the referenced method by (1) renaming, removing, or replacing $M$, or (2) changing the parameter list of $M$. Similarly, among the conflicts related to other entities, most conflicts occurred because of the references to missing entities. 
%because one branch added a reference to an entity, which entity was removed, replaced, or renamed by another branch (as demonstrated in Figure~\ref{fig:conflicts} (b)). 
Such conflicts usually produce the compilation errors of unresolvable referenced entities. 

\begin{figure}[h]
\centering
\includegraphics[width=0.43\textwidth]{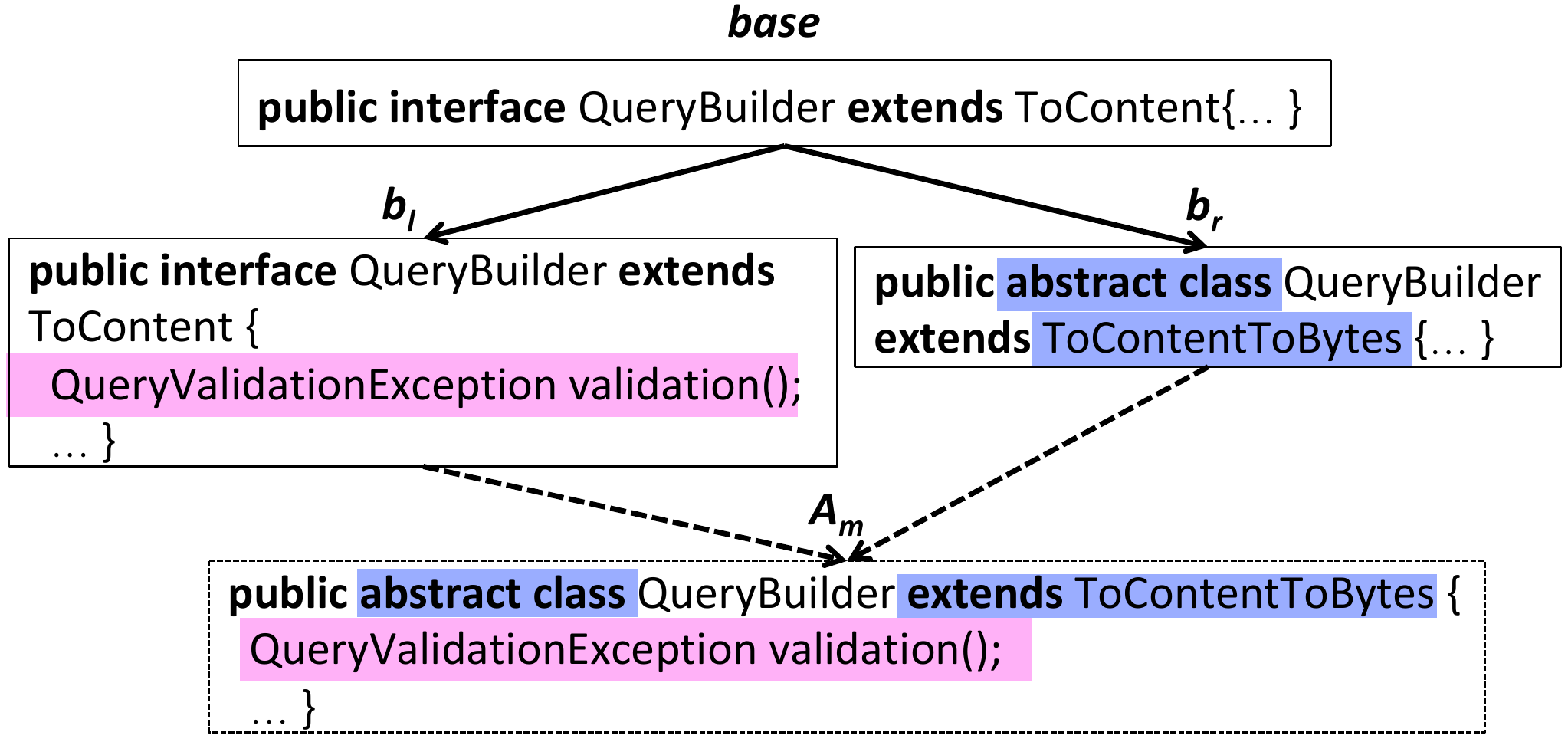}
%\vspace{-1.5em}
\caption{A compiling conflict that produces an errorneous abstract class definition}
\label{fig:scIntro-example}
\end{figure}
%\vspace{-2em}

In addition to the typical broken def-use relationship of entities mentioned above, some compiling conflicts can produce invalid entity definition. As illustrated by Figure~\ref{fig:scIntro-example}, $b_l$ declares a new method \codefont{validation()} in an existing interface \codefont{QueryBuilder}, while $b_r$ converts the interface to an abstract class. Although the edits can be textually integrated, the resulting software $A_m$ has an errorneous
abstract class declaration. Java compilers usually prompt developers to fix such errors by either adding method body for \codefont{validation()} or declaring it as an abstract method (i.e., using the annotation \codefont{@abstract}).

\begin{tcolorbox}
	\textbf{Finding 2:} \emph{
	93\% of compiling conflicts were related to either Java classes, methods, or fields.  91\% of conflicts were introduced because the conflicting edits broke the referencer-referencee relationship of entities. 
}
\end{tcolorbox}

\subsubsection{Characterization of Dynamic Conflicts}
\label{sec:s2cIntro}
We identified two reasons to explain the four inspected dynamic conflicts. 

\begin{figure}[h]
\centering
\includegraphics[width=0.49\textwidth]{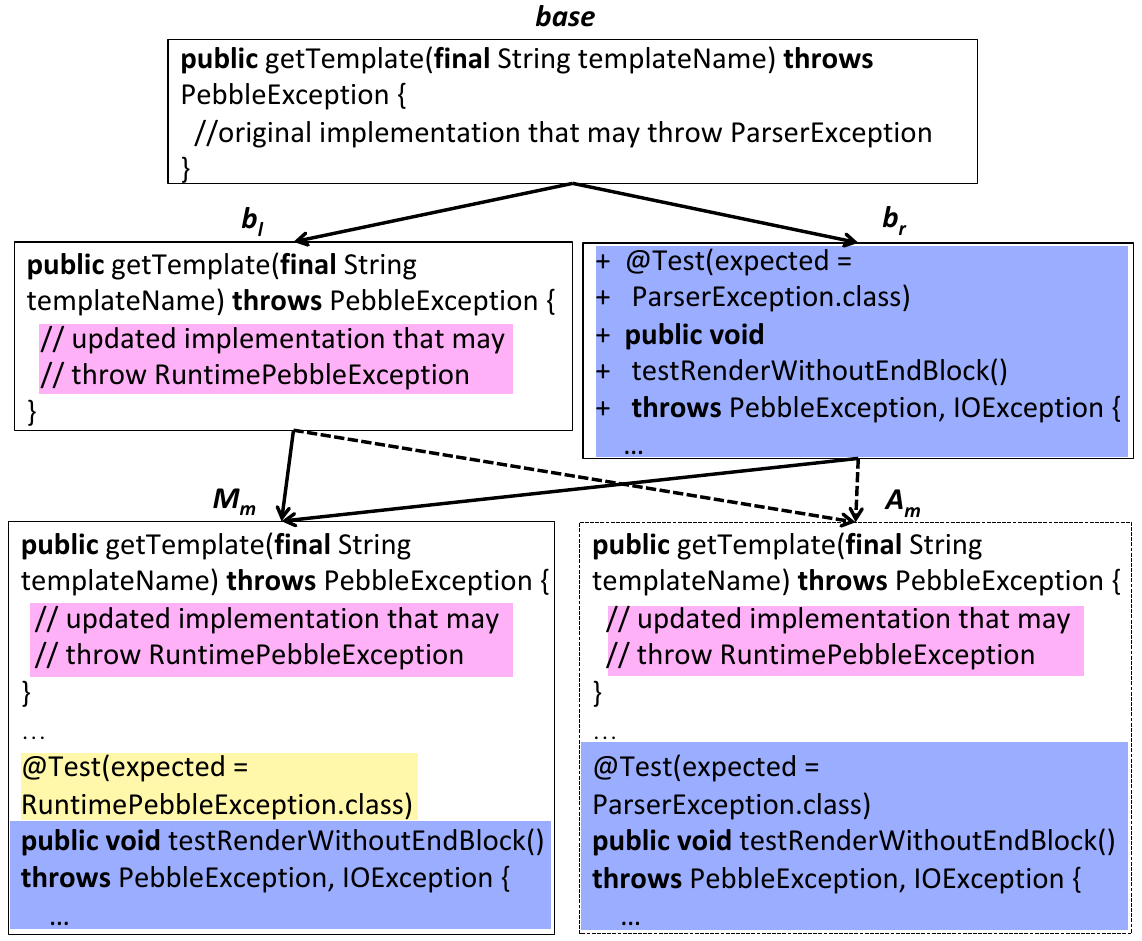}
%\vspace{-0.5em}
\caption{A dynamic conflict that causes incorrect test oracle}
\label{fig:s2cIntro-example}
\end{figure}

\noindent
\textbf{1. Incorrect Test Oracle}. Three conflicts occurred for this reason. 
As shown in Figure~\ref{fig:s2cIntro-example}, $b_l$ replaces the implementation of method \codefont{getTemplate(...)} such that a new exception \codefont{RuntimePebbleException} is thrown. Meanwhile, $b_r$ adds some test cases, assuming that \codefont{getTemplate(...)} is unchanged and still throws the original exception \codefont{ParserException}. 
Consequently, these test cases fail because the expected exception type in the test oracle (i.e., \codefont{ParserException}) does not match the actual thrown exceptions (i.e., \codefont{RuntimePebbleException}).

\begin{figure}
\centering
\includegraphics[width=0.43\textwidth]{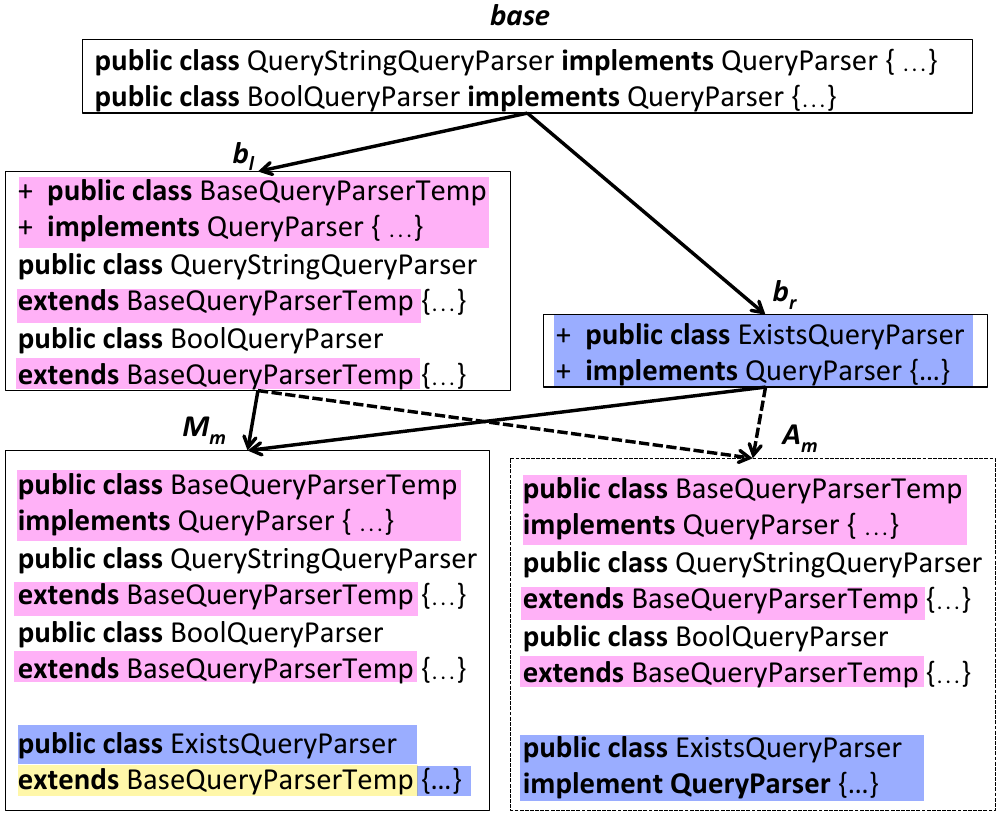}
\vspace{-0.5em}
\caption{A dynamic conflict that incorrectly adds a class}
\label{fig:s2cIntro-example2}
\end{figure}

\noindent
\textbf{2. Incorrectly Added Class Declaration.} 
One manually detected conflict happened for this reason. 
As shown in Figure~\ref{fig:s2cIntro-example2}, $b_l$ refactors the class hierarchy such that the interface \codefont{QueryParser} is implemented by only one class---the new class \codefont{BaseQueryParserTemp}, and all other classes originally implementing the interface are changed to instead inherit from this new class. 
On the other hand, $b_r$ defines a new class \codefont{ExistsQueryParser} to implement the original interface. 
Blindly merging the two versions can break the new class hierarchy design that $b_l$ tries to realize. Therefore, in $M$, developers modified \codefont{ExistsQueryParser} to instead inherit from \codefont{BaseQueryParserTemp}. 
%the original branches could not be automatically merged.
% due to textual conflicts. 

\begin{tcolorbox}
	\textbf{Finding 3:} \emph{
	The observed dynamic conflicts were introduced for two reasons: (1) the inconsistent changes between code implementation and test oracle, and (2) the inconsistent maintenance of the same class hierarchy.
}
\end{tcolorbox}

\vspace{-0.5em}
\subsection{RQ2: How Did Developers Manually Resolve Conflicts?}
\label{sec:rq2}

Generally speaking, developers fixed conflicts by either keeping the edits from $b_l$ (\textbf{L}), keeping the edits from $b_r$ (\textbf{R}), keeping part or all edits from both branches (\textbf{L + R}), or applying extra edits after merging the branches (\textbf{L + R + M}). 
%However, depending on the conflict types, developers' resolution strategies vary a lot. 

\subsubsection{Resolutions for Textual Conflicts}
\label{sec:tcResolve}

As shown in Table~\ref{tab:tcResolve}, developers resolved 69\% of textual conflicts by taking the edits from $b_l$ or $b_r$.  
%Most of these conflicts belong to 
%Especially, most conflicts of  \textbf{update-update}, \textbf{delete-update}, and \textbf{insert-insert} were resolved in such ways. 
There is no evidence showing that developers usually preferred one branch over the other, probably because developers' decision-making often depends on the applied edits and edit locations. 
%and their context. 
Interestingly, for all six inspected \textbf{update-move} conflicts, developers always tried to integrate the branches instead of inheriting from one branch. 
We further grouped conflicts based on (1) their types and (2) the commits from which they were extracted. 
Among the 19 groups of conflicts identified in this way, 15 groups were resolved consistently. 
Namely, for the multiple conflicts in each of these 15 groups, developers took the same resolution strategy: 7 groups were resolved via \textbf{L}, 4 groups were resolved via \textbf{R}, 1 group were resolved via \textbf{L+R}, and 3 groups were resolved via \textbf{L + R + M}. %It implies that developers consistently resolved the same-typed conflicts within the same commit. 

\begin{table}[h]
\caption{Resolutions for different kinds of textual conflicts}
\label{tab:tcResolve}
\vspace{-0.5em}
\centering
\footnotesize
\begin{tabular}{l|rrrr}
\toprule
\textbf{Conflict Type} & \textbf{L} & \textbf{R} & \textbf{L + R} & \textbf{L + R + M} \\ \hline
\toprule
\textbf{update-update} & 17 & 13 & 13 & 8\\ \hline
\textbf{delete-update} & 14 & 9 & - & 2\\ \hline
\textbf{insert-insert} & 5 & 9 & 1 & -\\ \hline
\textbf{update-move} & - & - & 2 & 4\\ \hline
\textbf{delete-move} & 2 & - & - & - \\ \hline
\textbf{move-move} & -& - &1 & -\\ \bottomrule
\textbf{Total} & 38 & 31 & 17 & 14 \\ \bottomrule
\multicolumn{5}{l}{``-'' indicates ``zero-entry''}
\end{tabular}
\end{table}

\begin{tcolorbox}
	\textbf{Finding 4:} \emph{
	Developers handled 69\% of textual conflicts by giving up the edits in one branch; they fixed the other 31\% of conflicts by somehow integrating the edits from both sides.  
}
\end{tcolorbox}

\subsubsection{Resolutions for Compiling Conflicts}
Although there are many conflict types listed in Table~\ref{tab:scIntro},
to simplify discussion, we merged these types into three major categories, as shown in Table~\ref{tab:scResolve}. 

\begin{table}[h]
\caption{Resolutions for compiling conflicts}
\label{tab:scResolve}
\vspace{-0.5em}
\centering
\footnotesize
\begin{tabular}{l|rrr}
\toprule
\textbf{Category} & \textbf{L} & \textbf{R} & \textbf{L + R + M} \\ \hline
\toprule
\textbf{broken def-use} & 4 & 1 & 86\\ \hline
\textbf{invalid def} & - & - & 6 \\ \hline
\textbf{other} & - & - & 3 \\ \bottomrule
\textbf{Total} & 4& 1 & 95  \\ \bottomrule
\multicolumn{4}{l}{``-'' indicates ``zero-entry''} 

\end{tabular}
\end{table}

\begin{itemize}
\item
\textbf{Broken def-use} includes all conflicts for which an added entity usage (e.g., method call, field access, or class reference) refers to a missing entity definition. This category corresponds to the finer-grained types in Table~\ref{tab:scIntro} that match the pattern ``Referencing a missing entity *''. 
\item 
\textbf{Invalid def} includes the cases where two branches declare or update the definition of related entities, creating problematic entity definitions. 
This category covers the following types listed in Table~\ref{tab:scIntro}: ``Restructuring an interface $I$'', ``Updating the parameter list of $M$'', ``Restructuring a method $M$'', and ``Adding duplicated declarations for variable $V$''. 
\item
\textbf{Other} covers the cases not included by the above categories.
%the above-mentioned two categories.%such as ``Importing a missing class $C$'' and ``Invoking a method $M$ with an updated return type''. 
\end{itemize}

By definition, developers cannot resolve compiling conflicts by na{\"i}vely integrating the edits from both sides (i.e., \textbf{L+R}), because such integration can cause compilation errors. Therefore, in Table~\ref{tab:scResolve}, we only list three resolution strategies: \textbf{L}, \textbf{R}, and \textbf{L + R + M}. According to the table, developers resolved 95 conflicts by applying extra edits after merging branches. 
It implies that developers usually integrated as many edits as possible, as long as those edits could be orchestrated with moderate effort. 

\begin{figure}
\includegraphics[width=0.49\textwidth]{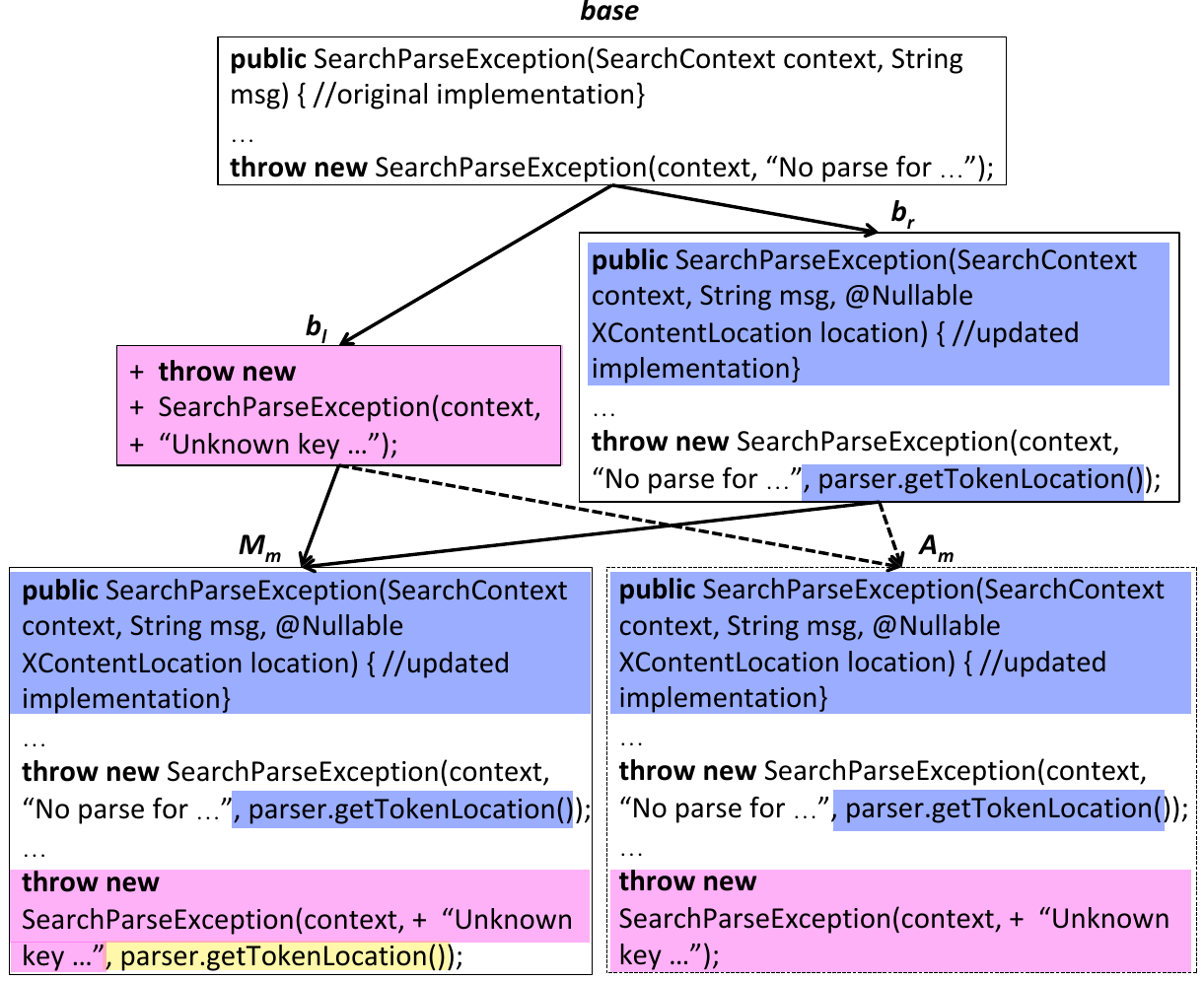}
\caption{A compiling conflict that developers resolved by consistently updating method calls}
\label{fig:scResolve-example}
\end{figure}

Figure~\ref{fig:scResolve-example} shows a typical way that developers took to resolve conflicts. In this example, $b_l$ inserts a method call to \codefont{SearchParse\-Exception(...)}; $b_r$ updates the method signature by adding the parameter \codefont{location}, and consistently modifies the method calls by passing in one more value \codefont{parser.getTokenLocation()}. 
Blindly combining these edits can produce a broken def-use chain between the declared method and the added method call. 
Therefore, $M_m$ contains developers' extra edit to similarly update the new method invocation.  
Although the extra edits in $M_m$ can be more complex for many conflicts (e.g., inserting, deleting, or moving statements),
%Actually, the adaptive edits that developers introduced to $M_m$ were not always so simple; developers might 
%introduce complex edits by also inserting, deleting, or moving statements. 
%What is more important, 
%However, 
%Interestingly, we found that 
%No matter what the edit content is, 
%the adaptive edits developers applied were usually similar to some of the edits in one branch. 
such edits were usually similar to some of the edits in one branch.

Additionally, developers resolved five other conflicts by keeping $b_l$ or $b_r$. Four of these conflicts are about adding a reference to a class whose import declaration is removed by the other branch. Naturally, developers resolved the conflicts by keeping the branches with imports.
%by inheriting from the versions where the imports still exist. 
Another conflict is about adding a reference to a variable whose declaration is removed by the other branch. 
Similarly, developers kept the branch that has the original declaration. 

\begin{tcolorbox}
	\textbf{Finding 5:} \emph{
	Developers handled 95\% of compiling conflicts by applying extra edits to the integrated version. These extra edits were usually similar to some of the edits in one branch.%Different from what they usually did for textual conflicts, developers kept the edits from both branches in the merged version. 
}
\end{tcolorbox}

\subsubsection{Resolutions for Dynamic Conflicts}
All dynamic conflicts were resolved with the strategy \textbf{L + R + M}. For the three conflicts with incorrect test oracles, developers applied extra edits to correct those oracles. In particular, the corrective edits in $M_m$ for two conflicts were similar to those applied in one branch. However, the corrective edits for the third conflict is different from those in both branches. As shown in Figure~\ref{fig:s2cIntro-example}, the corrective edit in $M_m$ is:
\begin{lstlisting}
-  @Test(expected = ParserException.class)
+  @Test(expected = RuntimePebbleException.class)
\end{lstlisting}
Meanwhile, the adaptive edit consistently applied in $b_l$ is:
\begin{lstlisting}
-  @Test(expected = ParserException.class)
+  @Test
\end{lstlisting}
Theoretically, even if developers replicate their edits in $b_l$ to $M_m$, they can still fix the conflict. This example still supports our observation that 
the extra edits in $M_m$ are usually similar to those in one branch. 
Finally, 
for the conflict producing incorrectly added class declaration, developers resolved the problem by similarly applying their $b_l$ edits to $M_m$ (see Figure~\ref{fig:s2cIntro-example2}).

\begin{tcolorbox}
	\textbf{Finding 6:} \emph{
	Developers handled all dynamic conflicts by applying extra edits to the integrated versions; 75\% of these extra edits were similar to those from one branch. 
}
\end{tcolorbox}

\vspace{-0.5em}

\subsection{RQ3: What Conflicts Cannot Be Handled by Current Tools?}
\label{sec:rq3}
%As mentioned in Section~\ref{sec:approach}, we revealed 100 textual conflicts that cannot be automatically resolved by either git-merge or JDime. 
This section discusses the capability of current tools in two aspects: conflict detection (Section~\ref{sec:cdetect}) and resolution (Section~\ref{sec:cresolve}). 

\subsubsection{Conflict Detection Capability of Existing Tools}
\label{sec:cdetect}

As shown in Table~\ref{tab:cdetect}, 
all explored tools could reveal the textual conflicts in our data set. However, only Crystal-like merge revealed 21\% of the inspected compiling conflicts and 25\% of dynamic conflicts. In particular, although we manually found most compiling conflicts in automatically unmergeable branches, Crystal-like merge cannot detect these conflicts because automatic compilation is feasible only when $A_m$ is generated. 
%check for compiling errors in unmerged software branches, neither can Crystal-like merge do that. 
Therefore, to more efficiently detect compiling conflicts, 
future tools can (1) statically compare the program structures of software branches, and (2) mimic compilers to identify any conflicting dependency between program entities (e.g., an added method call refers to a removed method declaration).

\begin{table}[h]
\centering
\footnotesize
\caption{Conflicts detected by existing tools}
\label{tab:cdetect}
\vspace{-0.5em}
\begin{tabular}{L{3cm}|R{1.2cm}|R{1.2cm} |R{1.2cm}}
\toprule
\textbf{Tools} & \textbf{Textual Conflicts} & \textbf{Compiling Conflicts} & \textbf{Dynamic Conflicts}\\ 
\toprule
\textbf{Text-based merge (e.g., git-merge)} & 100\% & 0\% & 0\% \\  \hline
\textbf{JDime} & 100\% & 0\% & 0\%\\ \hline
\textbf{Crystal-like merge (e.g., Crystal and WeCode~\cite{Guimaraes:2012})} & 100\% & 21\% & 25\% \\
%\textbf{SafeMerge~\cite{sousa2018verified}} & \xmark & \xmark & \xmark \\
\bottomrule
\end{tabular}
\end{table}

Similarly, Crystal-like approaches are not effective to reveal dynamic conflicts, either. 
 Two reasons can explain such deficiency. First, automatic testing does not help when no $A_m$ is generated or executable. Second, when dynamic conflicts are not covered by any test run, testing cannot reveal such conflicts. 
 %even though such conflicts can cause runtime errors, testing cannot trigger those errors. 
 Sousa et al.~recently proposed SafeMerge---a verification algorithm to reason about the semantics of $base$, $b_l$, $b_r$, and $A_m$~\cite{sousa2018verified}. 
 By identifying any semantic equivalence between the four versions, the researchers showed that SafeMerge correctly revealed three dynamic conflicts without running any test.
 However, SafeMerge verifies one procedure at a time, assuming that other procedures never affect this one; the approach cannot detect conflicts when the assumption does not hold.  
 %requires that the edits by $b_l$ and $b_r$ must be in the same Java method; 
 %branches simultaneously edit different Java methods or files. 
In our data set, unfortunately, SafeMerge cannot detect any of the studied dynamic conflicts because the edits applied in distinct program entities affect each other.

 % because the conflicting edits were all applied in distinct program entities. 
 
\begin{tcolorbox}
	\textbf{Finding 7:} \emph{In our data set, 79\% of compiling conflicts and 75\% of dynamic  conflicts could NOT be revealed by any existing tool, meaning that we still need better detection tools that require for no compilers or test execution. }
\end{tcolorbox}

\subsubsection{Conflict Resolution Capability of Existing Tools}
\label{sec:cresolve}
Due to the approach we took to collect data, none of the tools mentioned in Section~\ref{sec:cdetect} could resolve the studied conflicts. 

Additionally, AutoMerge is a tool that recommends alternative resolutions for developers' consideration~\cite{Zhu:2018}. 
For instance, suppose that $base$ has two statements $(S_1;S_2)$, which were updated to $(S_1';S_2')$ by $b_l$ and updated to $(S_1'';S_2'')$ by $b_r$ ($S_1'\neq S_1'',S_2'\neq S_2''$). AutoMerge enumerates all possible combinations between the updated statements, e.g., $(S_1';S_2'')$, to suggest candidate resolutions. 
In our data set, the actual resolutions for 86\% of textual conflicts can be covered by such candidate resolutions, meaning that AutoMerge can potentially resolve 86\% of the inspected textual conflicts. However, given a conflict, AutoMerge cannot automatically decide for developers which candidate resolution to take. 

WeCode resolves certain kinds of textual conflicts in predefined ways~\cite{Guimaraes:2012}. Specifically, for the \textbf{update-update} conflicts where the same variable is assigned to different constant integers, WeCode simply assigns the variable to a predefined default value (e.g., 0). 
For \textbf{delete-update} conflicts, WeCode keeps the updated statement. 
However, WeCode does not ensure the correctness of its resolutions.
Theoretically, WeCode can resolve at most 30 conflicts 
(i.e. 5 \textbf{update-upate} for variable assignments + 25 \textbf{delete-update}).
% in our data set. 

Xing et al.~built a semi-automatic approach to resolve the dynamic conflicts implied by test failures~\cite{Xing2019}. 
After test failures, developers are required to define extra tests to show the expected program behaviors; an off-the-shelf Automatic Program Repair (APR) system (i.e., kGenProg~\cite{Higo2018}) is then used to generate patches such that the patched program can pass all tests. Theoretically, this approach can resolve only one conflict in our data set---the conflict that passes compilation but fails testing. In reality, the approach is limited by two assumptions. First, the newly defined tests can cover all desired program behaviors of the merged software. Second, APR can create the correct patch within a given period of time. 

\begin{tcolorbox}
	\textbf{Finding 8:} \emph{Theoretically, the resolutions for 86\% of textual conflicts and 25\% of dynamic conflicts can be generated by existing tools. However, these tools do not guarantee to automatically choose the correct resolutions over the other generated candidates.
		No tool resolves compiling conflicts. 
%	In our data set, 79\% of compiling conflicts and 75\% of dynamic  conflicts could not be revealed by any existing tool, meaning that we still need better detection tools that require for no compilers or test execution. 
}
\end{tcolorbox}

\vspace{-0.5em}
\section{Our Recommendations}
\label{sec:discussion}

Our work reveals the characteristics of various conflicts and the resolution strategies of developers. These findings lead us to give the following recommendations on future tool design.

\paragraph{\textbf{Resolution Prediction for Textual Conflicts.}} 
Given a textual conflict, existing tools at most enumerate and suggest the feasible resolution alternatives, but cannot predict which suggestion developers finally take. According to Section~\ref{sec:tcResolve}, for the same-typed conflicts from the same commit, developers usually took the same resolution strategy. 
Inspired by this observation, we can build future tools that (1) monitor developers' resolutions (e.g., \textbf{L}) for certain types of conflicts (e.g., \textbf{update-update}),  and (2) dynamically suggest the same resolution strategies for the same-typed unresolved conflicts occurring in the same commit. In this way, these tools can minimize the manual effort required to fix textual conflicts, and thus help developers create the merged software much faster. 

\paragraph{\textbf{Detection of Higher-Order Conflicts.}}
%As mentioned in Section~\ref{sec:cdetect}, 
Existing tools detect higher-order conflicts 
via compiling and testing the $A_m$ created by text-based merge.
However, we observed that (1) many higher-order conflicts were \emph{actually} introduced by text-based merge; (2) dynamic conflicts do not always trigger test failures; and (3) not every test failure can help pinpoint the related conflict. 
%and dynamic conflicts do not always trigger test failures. 
%Second, semantic conflicts seldom triggered test failures, and even worse, the triggered test failures could not effectively locate the conflicting edits. 
All our observations motivate better approaches that conduct static program analysis to detect all types of conflicts at once. With a holistic view of the conflicts between branches, developers can prioritize conflicts based on their importance instead of the exposure sequence, and avoid introducing higher-order conflicts when handling textual conflicts. 

\paragraph{\textbf{Resolution of Higher-Order Conflicts.}}
Current tools rarely fix higher-order conflicts. 
However, we observed that developers typically took the \textbf{L + R + M} strategy by applying extra edits to $A_m$, which edits were similar to those applied in one branch. Some systematic editing tools like SYDIT~\cite{Meng:2011} and LASE~\cite{Meng2013:lase} can generalize abstract program transformations from concrete code change examples, and repetitively apply similar transformations to similar code snippets. 
It is promising to extend these tools for automatic conflict resolution. 
For instance, given a code snippet with a higher-order conflict, we can extend LASE to search for similar code that was edited in one branch, and then similarly apply the edit to $A_m$.

\vspace{-0.5em}
\section{Related Work}
The related work includes empirical studies of merge conflicts, automatic  merge approaches, and change recommendation systems.  

\vspace{-0.5em}
\subsection{Empirical Studies on Merge Conflicts}
Several studies were recently conducted to characterize merge conflicts~\cite{Estler2014,Ahmed2017,Lebetaenich:2018,Mahmoudi2019,Owhadi-Kareshk2019}. 
Specifically, 
Le{$\beta$}enich et al.~surveyed 41 developers and identified 7 potential indicators (e.g., \# of changed files in both branches) for merge conflicts~\cite{Lebetaenich:2018}.
With a further empirical study of the indicators, the researchers found that none of the indicators can predict the frequency of conflicts. 
Similarly, Owhadi-Kareshk et al.~defined 9 features to characterize merging scenarios, and trained a machine learning model that predicted conflicts with 57\%-68\% accuracy~\cite{Owhadi-Kareshk2019}. 
Mahmoudi et al.~observed that certain refactoring types (e.g., Extract Method) tend to be more problematic for merge conflicts~\cite{Mahmoudi2019}.
Ahmed et al.~showed that smelly entities are three times more likely to be involved in conflicts~\cite{Ahmed2017}.
Estler et al.~analyzed the relationship between merge conflicts and developers' awareness, showing that lack of awareness occurs more frequently than conflicts~\cite{Estler2014}. 
All these studies only examined the relationship between textual conflicts and other code features (e.g., smell) or developers' coding activities (e.g., refactoring). 
%However, they only examined the textual conflicts reported by text-based merge, without exploring other kinds of conflicts. 
%how conflicts are resolved or should be resolved automatically.

Fewer studies characterized conflicts themselves~\cite{Yuzuki2015,Nguyen2018,Nelson2018}. 
For instance, 
Yuzuki et al.~inspected hundreds of textual conflicts~\cite{Yuzuki2015}. They observed that 44\% of conflicts were caused by conflicting updates on the same line of code, and developers resolved 99\% of conflicts by taking either the left- or right- version of code.
Our study revealed similar findings. 
Nguyen et al.~observed that (1) a higher integration rate of a project does not generate a higher unresolved conflict rate, and (2) developers are more likely to 
take the left- or right- version of code to resolve higher-order conflicts~\cite{Nguyen2018}.
Nelson et al.~conducted both surveys and interviews with developers~\cite{Nelson2018}. They learnt that developers deferred responding to conflicts based on their perception of the complexity of the conflicting code, and that deferring affects the workflow of the entire team. 

Our study is different from prior work in two aspects. First, 
by experimenting with various conflict detection tools, 
we explored the gap between hard-to-detect conflicts and capabilities of current tools, and suggested future research to close the gap. 
Second, by 
investigating the resolution strategies adopted by developers, we summarized the fixing patterns for conflicts and identified opportunities for further automation. 
%and characterized the limit of automatic approaches.
%characterized the limit of current and future automatic tools. 

\vspace{-0.5em}
\subsection{Automatic Software Merge Approaches}
Tools were built to detect or resolve merge conflicts~\cite{Brun:2011,Apel:2011,Apel:2012,Guimaraes:2012,Brun2013,Leenich2014,Zhu:2018,Xing2019}.
For instance, %semistructured merge inherits the generality of unstructured merge and expressiveness of structured merge~\cite{Apel:2011}. In particular, the implemented tool---
FSTMerger combines structured and unstructured merge~\cite{Apel:2011}. It
matches Java methods purely based on the method signatures, and integrates the content of matched methods via text-based merge. 
WeCode continuously merges the committed and uncommitted changes in software branches, and raises developers' awareness when any conflict is detected~\cite{Guimaraes:2012}. The system detects conflicts matching certain patterns and resolves conflicts in predefined ways. 
Xing et al.~proposed a semi-automatic approach to resolve dynamic conflicts~\cite{Xing2019}. When a merged program fails one or more tests, the approach requires developers to define more tests, and then adopts
%developers are required to define more tests to demonstrate the program's expected behaviors. Then the approach leverages 
Automatic Program Repair (APR) to generate candidate fixes to resolve conflicts until all tests are passed. 
%AutoMerge represents textually conflicting branches with Version Space Algebra (VSA)~\cite{Zhu:2018}. Intuitively, each pair of conflicting statements are represented as alternative child nodes of their structural parent node. 
%By enumerating all possible combinations between the alternatives under different parent nodes, AutoMerge recommends multiple resolutions to developers. 

We took both manual and automatic approaches to reveal a spectrum of conflicts. By revealing the conflicts overlooked by existing tools, we suggested new tools for conflict detection. By characterizing developers' strategies to resolve conflicts, we checked whether existing tools mimic humans' resolutions, and revealed some unknown but frequently adopted strategies that are automatable.

\vspace{-0.5em}
\subsection{Change Recommendation Systems}
Based on the insight that similar code is likely to be changed similarly, researchers proposed tools to recommend code changes or facilitate systematic program editing~\cite{Miller2001,Li2004,Jiang2007:CDC,Nguyen2009:clever,Nguyen2010:libsync,Meng:2011,Meng2013:lase}. 
With more details, simultaneous editing enables developers to simultaneously edit multiple pre-selected code fragments in the same way~\cite{Miller2001}. While a developer interactively demonstrates the edit operations in one fragment, the tool replicates the lexical edits (e.g., copy a line) to other fragments. 
CP-Miner identifies code clones (i.e., similar code snippets), and detects copy-paste related bugs if clones have inconsistent identifier or context mappings~\cite{Li2004}.
Given two or more similarly changed code examples, LASE extracts the common edit operations, infers a general program transformation, and leverages the transformation to locate code for similar edits~\cite{Meng2013:lase}. 

Although both software merge and change recommendation systems are active research areas, our study is the first piece of work that identifies a nice connection between the areas. 
Our study 
uncovers various scenarios where developers repetitively applied similar or related edits to resolve conflicts. It enlightens future research to improve change recommendations such that the conflict-specific systematic edits can be dynamically generated and applied.

\vspace{-0.5em}
\section{Threats to Validity}

\paragraph{Threats to External Validity}
Our study is based on the 204 conflicts extracted from 15 Java project repositories.
The characteristics these conflicts present and the observed resolutions may not generalize well to other conflicts, other projects, or other programming languages. In the future, we plan to reduce this threat by considering the projects of other programming languages, investigating more projects, and including more samples in our data set. 
%limited number of projects, people, expertise

\paragraph{Threats to Construct Validity}
We took a hybrid approach to detect conflicts. In particular, as automatic tools only revealed 21 compiling conflicts and 1 dynamic conflict, we manually analyzed unmerged software branches to detect more higher-order conflicts. 
It is possible that certain conflicts are easier to manually detect than others, so the collected data can be subject to human bias. To overcome this limitation, we leveraged cross-validation (i.e., having multiple people to examine the same conflicts) to reduce random errors and avoid bias. In the future, we will build better tools to detect various conflicts more systematically and efficiently.  

\vspace{-0.5em}
\section{Conclusion}

Prior empirical studies showed that merge conflicts frequently occur and solving conflicts is important but challenging. In this study, we comprehensively studied all kinds of conflicts and their resolutions, and characterized the conflicts that cannot be handled by existing tools. Different from prior studies that mainly focus on textual conflicts, 
our study is wider and deeper for two reasons. First, we also examined higher-order conflicts, which are harder to reveal and resolve. Second, 
by assessing related tools with the same data set, 
%by putting related tools in the same context, 
we evaluated the capabilities of current approaches in a theoretical way and characterized the limit of current approach design. 

Our study provides multiple insights. First, same-typed textual conflicts in the same commits were usually resolved with the same strategy (i.e., \textbf{L}, \textbf{R}, \textbf{L + R}, or \textbf{L + R + M}). Therefore, even though it is almost impossible to predict developers' strategy for any arbitrary textual conflict, such prediction is feasible given the resolution of other textual conflicts in the same merging commit.
Second, text-based merge can produce higher-order conflicts by silently integrating 
semantically conflicting edits, while compilation and testing usually fail to capture  the produced conflicts. It means that better tools are desperately needed to detect all kinds of conflicts altogether, instead of detecting certain conflicts at the cost of introducing other conflicts. 
Third, developers usually resolved higher-order conflicts by consistently applying similar edits to similar code locations. By automating such practices, future tools can resolve many higher-order conflicts that we observed. Our future work is on building the (semi)-automatic tools enlightened by this study.

\bibliographystyle{abbrv}
\bibliography{bowen}

\end{document}